\documentclass[
reprint,
superscriptaddress,
 amsmath,amssymb,
 aps,
prapplied,
twocolumns, 
]{revtex4-2}

\usepackage{graphicx}
\usepackage{dcolumn}
\usepackage{bm}
\usepackage{mhchem}
\usepackage{placeins} 

\graphicspath{ {Main_figures/} }

\begin{document}


\title{Local and Non-local Microwave Impedance of a Three-Terminal Hybrid Device}

\author{B.~Harlech-Jones$^*$}
\affiliation{ARC Centre of Excellence for Engineered Quantum Systems, School of Physics, The University of Sydney, Sydney, NSW 2006, Australia}

\author{S.~J.~Waddy$^*$}
\affiliation{ARC Centre of Excellence for Engineered Quantum Systems, School of Physics, The University of Sydney, Sydney, NSW 2006, Australia}

\author{J.~D.~S.~Witt} 
\affiliation{ARC Centre of Excellence for Engineered Quantum Systems, School of Physics, The University of Sydney, Sydney, NSW 2006, Australia}

\author{D.~Govender}
\affiliation{Microsoft Quantum Sydney, The University of Sydney, Sydney, NSW 2006, Australia} 

\author{L.~Casparis}
\affiliation{Microsoft Quantum Copenhagen, Kanalvej 7,
2800 Kongens, Lyngby, Denmark}

\author{E.~Martinez}
\affiliation{Microsoft Quantum Copenhagen, Kanalvej 7,
2800 Kongens, Lyngby, Denmark}

\author{R.~Kallaher}
\affiliation{Microsoft Quantum Purdue, Purdue University, West Lafayette, Indiana, USA}

\author{S.~Gronin}
\affiliation{Microsoft Quantum Purdue, Purdue University, West Lafayette, Indiana, USA}

\author{G.~Gardner}
\affiliation{Microsoft Quantum Purdue, Purdue University, West Lafayette, Indiana, USA}

\author{M.~J.~Manfra}
\affiliation{Microsoft Quantum Purdue, Purdue University, West Lafayette, Indiana, USA}
\affiliation{Department of Physics and Astronomy, Purdue University, West Lafayette, Indiana, USA}

\author{D.~J.~Reilly$^\dagger$}
\affiliation{Microsoft Quantum Sydney, The University of Sydney, Sydney, NSW 2006, Australia}
\affiliation{ARC Centre of Excellence for Engineered Quantum Systems, School of Physics, The University of Sydney, Sydney, NSW 2006, Australia}

\date{\today}

\begin{abstract}
We report microwave impedance measurements of a superconductor-semiconductor hybrid nanowire device with three terminals (3T). Our technique makes use of transmission line resonators to acquire the nine complex scattering matrix parameters ($S$-parameters) of the device on fast timescales and across a spectrum of frequencies spanning 0.3 - 7 GHz. Via comparison with dc-transport measurements, we examine the utility of this technique for probing the local and non-local response of 3T devices where capacitive and inductive contributions can play a role. Such measurements require careful interpretation but may be of use in discerning true Majorana zero modes from trivial states arising from disorder. 

\end{abstract}
\maketitle

Majorana zero modes (MZMs) are predicted to emerge from the twisted band structure that occurs when superconductivity is induced in a one-dimensional (1D) semiconductor nanowire with large spin-orbit coupling \cite{Lutchyn_2010,Oreg_2010}. The emergence of MZMs, bound to the ends of the wire, is a consequence of bulk-boundary correspondence, signalling the opening of an energy gap associated with a topological phase. These exotic states of matter are currently of interest for protecting quantum information from noise by encoding it non-locally in pairs of Majorana states that are spatially separated by the length of the wire \cite{Das_sarma_2015}.

It is natural then, that measurements to detect topological phases should also probe the system in a way that is sensitive to non-local properties \cite{Rosdahl2018}. Indeed, a means of identifying a bulk topological energy gap in hybrid superconductor-semiconductor devices has been developed recently, making use of local and non-local conductance measurements between spatially separated normal and superconducting leads in devices with three terminals \cite{Pikulin2021, TopoFound}. Taken together, local and non-local measurements can help distinguish true MZMs from topologically trivial states arising from disorder \cite{Frolov_21,TopoFound,Pikulin2021,pan2020}.

Here, we report complex microwave impedance measurements of a three-terminal (3T) superconductor-semiconductor hybrid device, incorporating both reflection and transmission measurements to simultaneously probe the local and non-local response on fast timescales. Our technique obtains the two nine-element scattering-parameter matrices $S_{ij}$ associated with the magnitude and microwave phase response for each of the permutations possible with the three ports of the device. Beyond improving the speed and sensitivity of 3T experiments, the microwave approach offers new qualitative information not accessible with low frequency conductance measurements \cite{Puglia2020,Frolov_21,TopoFound} by determining the reactive contribution to the impedance as a function of frequency. Detecting the quantum capacitance, for instance, can probe states that are isolated from transport paths by large tunnel barriers.  Examples may include `dark MZMs', where segments of the device are in a topological phase but, owing to disorder, are decoupled from the transport leads. Similarly, the presence of a supercurrent can lead to detectable (kinetic) inductance contributions to the reactance. Beyond the physics of topological devices, this measurement approach is likely of use more generally in high-frequency investigations of mesoscale structures with multiple terminals \cite{MultiT}.

\begin{figure*}
\includegraphics[width=\linewidth]{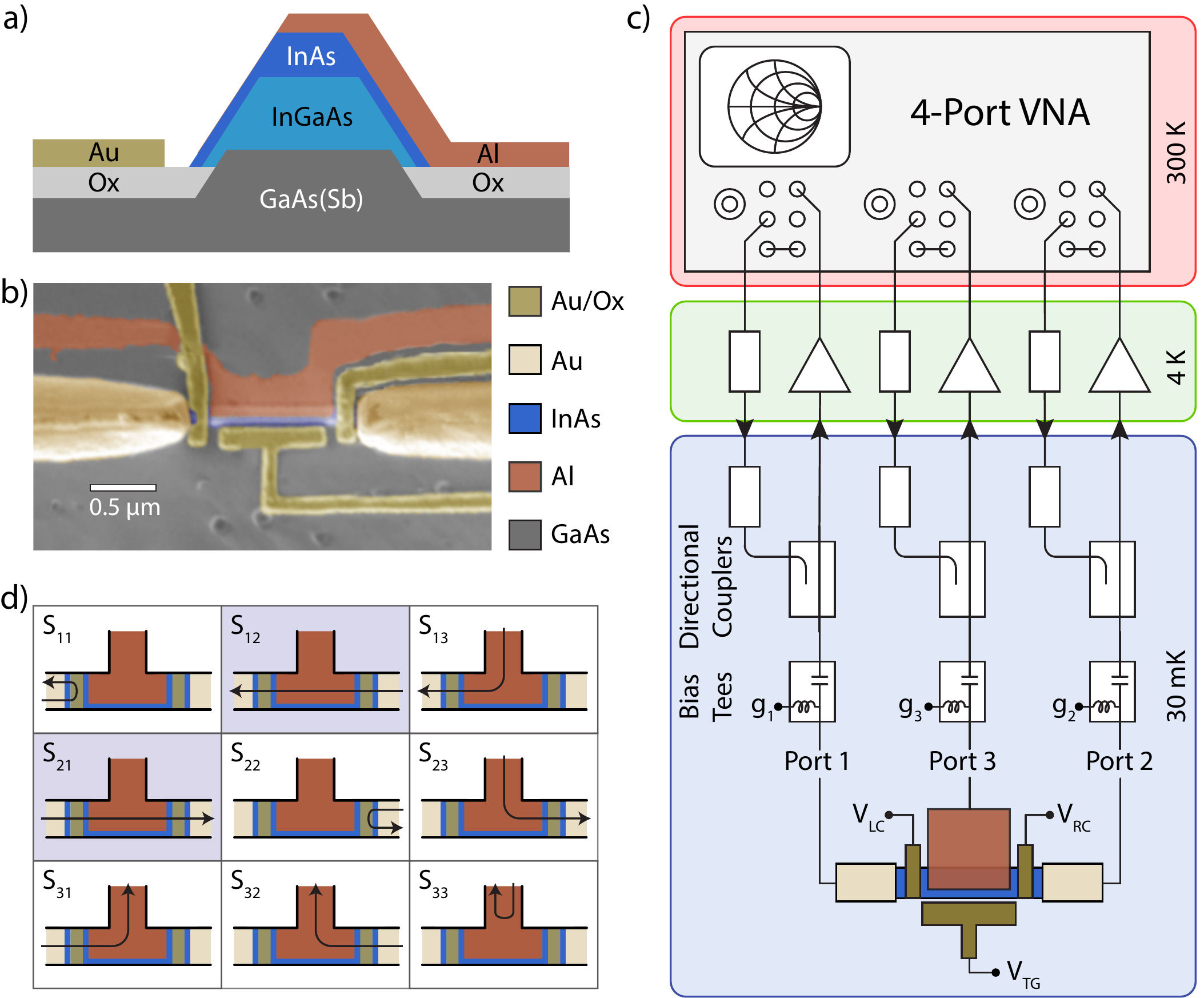}
\caption{\label{fig1} 
(a) Schematic cross-section through a nanowire device.
(b) False-colour SEM micrograph of the device showing cutter-gates, plunger gate, and 3 ports.
(c) Experimental setup: $S$-parameters were measured using a vector network analyzer (VNA), using the configurable test set option. The transmit (Tx) line was attenuated (boxes) at each temperature stage to reduce thermal noise and improve thermalization. Directional couplers are configured at the mixing chamber temperature stage. Conductance measurements via lock-in amplifiers and dc offset-bias were added to the microwave signal via bias-tees. Reflected and transmitted signals from each port are amplified using three cryogenic amplifiers that the 4~K stage of the device. (d) Measurement paths for the three terminal device. The `non-local' paths are highlighted in purple.}
\end{figure*}

\begin{figure*}
\includegraphics[width=\linewidth]{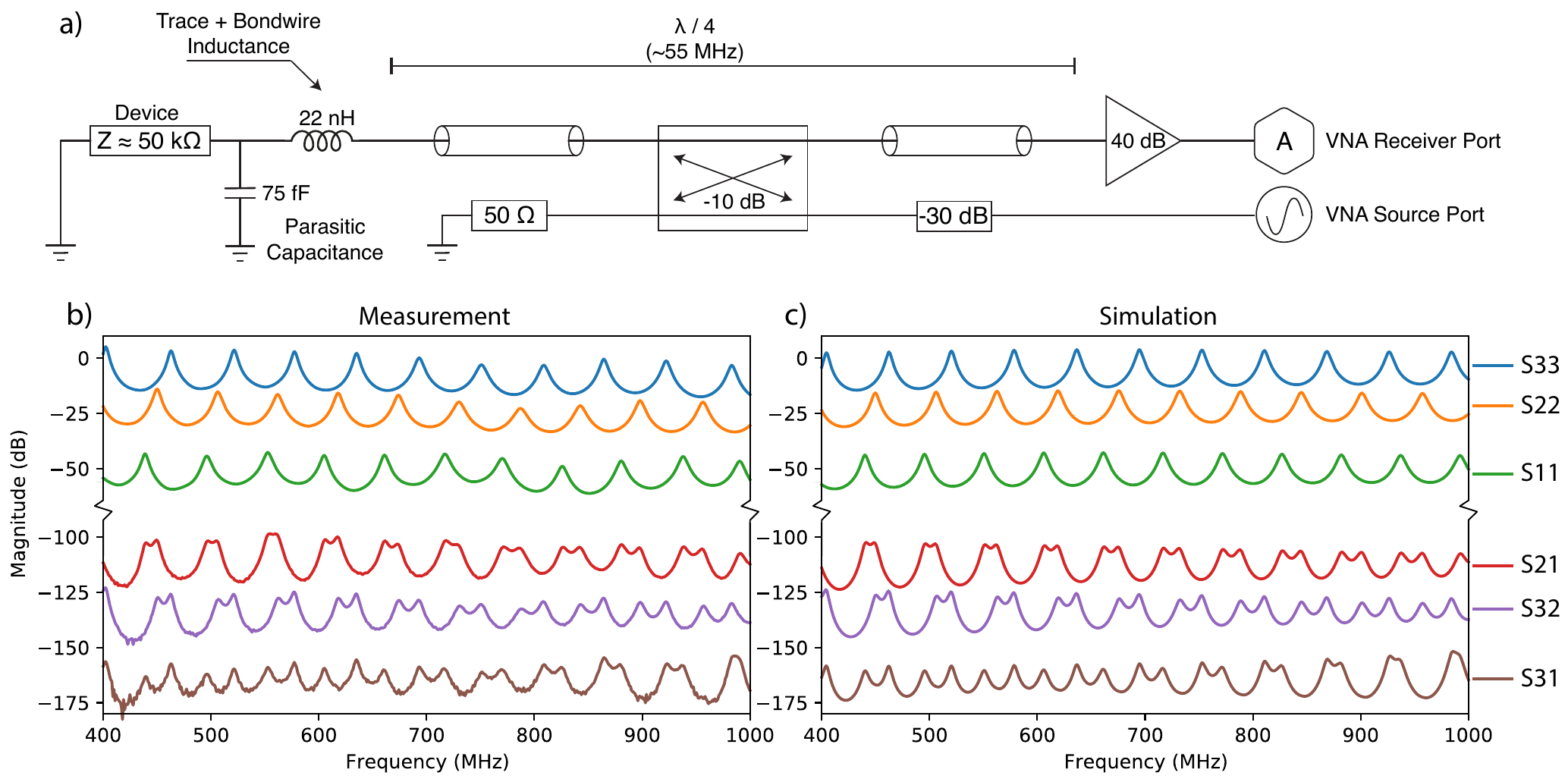}
\caption{\label{resonances} 
(a) Circuit diagram of the transmission line resonator (TLR) approach. We intentionally exploit the small impedance mismatch between the cryogenic amplifier and cable to create a standing wave that acts to transform the device impedance.
(b) Measured and (c) simulated resonant mode effects. Transmission windows occur when harmonic modes of each cable overlap. (For clarity, each successive trace is offset by -20~dB). }
\end{figure*} 

Our superconductor-semiconductor device is shown in Fig.~\ref{fig1}.(a) and (b) and comprises an epitaxial Al-InAs nanowire defined using selective area growth (SAG) via molecular beam epitaxy (MBE) on a GaAs substrate \cite{Krizek2018}. Aluminium is deposited in situ immediately after semiconductor growth to produce a clean interface \cite{Krugstrup2015, Krizek2018, Vaitiekenas2018} and selectively removed by a wet etch to leave only superconducting aluminium that proximitizes the wire. This Al layer also makes up the third (middle) superconducting terminal of the device \cite{Anselmetti2019}. Ohmic contacts comprising Ti-Au films are deposited on the ends of the nanowire and define each of the normal terminals. Finally, a dielectric layer of HfO$_2$ separates the Ti-Au top and cutter gates from the semiconductor. 
The device reported here has length of the proximitized region between the tunnel gates, $L\sim $1~$\mu$m. All measurements are made at the base temperature of a dilution refrigerator, $T<$~30~mK.

In a 3T device with gate-defined tunnel barriers, the `local' (differential) conductance probes tunneling between the left ($g_{13}$ = $di_{13}/dv_{13}$) or right ($g_{23}$ = $di_{23}/dv_{23}$) normal lead and the proximitized nanowire. Here the subscripts $ij$ refer to current that flows between the the transport terminals and voltage across them, [see Fig.~\ref{fig1}.(c)]. For bias energies $eV_{ij}$ below the superconducting gap, the middle superconducting terminal shunts current to ground over a length scale of order the coherence length, $\xi$. For long nanowires $L>\xi$, the presence of the grounded middle superconducting contact suppresses, to first-order, the `non-local' conductance between the normal terminals ($g_{12}$ = $di_{2}/dv_{1}$ and its converse $g_{21}$) and the two ends of the nanowire become decoupled. 

In contrast, for energies $eV_{ij}$ above the induced gap (but below the parent gap of aluminium), the non-local conductance becomes finite and its measurement provides a means of tracking the evolution of the gap with magnetic field and chemical potential (controlled by gates) \cite{Menard2020,Puglia2020}. Indeed, the closing and re-opening of the induced gap is a key signature of a topological phase transition in the bulk of the proximitized nanowire \cite{Rosdahl2018,TopoFound}. Taken together, a comparison of the local and non-local conductances \cite{Anselmetti2019}, and their evolution with field and gate bias, can provide a means of distinguishing MZMs from topologically trivial (Andreev) states \cite{Pikulin2021,TopoFound}. In this work we extend this picture to the microwave domain, where it is not obvious how the local and non-local transport signatures should manifest when the full complex impedance of the device is considered. Although in this work we do not probe exactly the parameter space where MZMs can manifest, our measurement approach carries over directly to that regime. %

In order to probe the local and non-local response of a 3T device at microwave frequencies it is necessary to measure both reflected and transmitted power from all three ports. Our setup for measuring the complex impedance of a 3T device is shown in Fig.~\ref{fig1}.(c). As well as the microwave response as a function of frequency, we also measure dc transport between each of the three terminals [labeled $g_1$, $g_2$, $g_3$ in Fig.~\ref{fig1}.(c)]. Transport is measured via a bias-tee that also enables the independent application of a voltage bias to all three contacts of the device [see supplemental for details]. To separate forward and reflected power the setup makes use of three directional couplers, mounted at the mixing chamber stage of the dilution refrigerator. Three cryogenic amplifiers mounted at the 4~K stage amplify returning signals that are then measured using a 4-port vector network analyzer with configurable test-set options. Taking all port  combinations of reflected and transmitted power we determine the 9-element scattering matrix ($S$-parameters), as shown schematically in Fig.~\ref{fig1}.(d). For each $S$-parameter it is also possible to separate the response in terms of magnitude and phase as a function of frequency. 

In contrast to standard rf-reflectometry which typically uses lumped-element resonators \cite{reflectreview}, our setup implements impedance matching between the device and the characteristic impedance of the system ($Z_0$=50~$\Omega$) using distributed transmission-line resonators (TLRs). These resonators are created using superconducting coaxial cables between the device and cryogenic amplifier, exploiting the impedance mismatch of the front-end of the amplifier to establish a standing wave that transforms the impedance of the device towards $Z_0$ at wavelengths corresponding to integer multiples of $\lambda/4$, i.e., every 55~MHz for a $Q$-factor of approximately 30. A circuit schematic of the TLR approach as well as measured and simulated $S$-parameter responses are shown in Fig.~\ref{resonances} [see supplemental for details]. 

The TLR produces a harmonic spectrum of resonance responses enabling impedance measurements at different frequencies, spanning a few 100~MHz to $>$~7~GHz. In our configuration, where all three ports are attached to TLRs, both the reflection and transmission response can be acquired, as is needed for local and non-local measurements. This is in contrast to lumped-element approaches to fast measurements, where the requirement that all resonators be matched to the same frequency may lead to crosstalk from shared parasitic elements. Alternatively,  using lumped element resonators with separate resonance frequencies for multi-port measurements is limited to reflection measurements ($S_{11}$) that probe only local responses.

\begin{figure}
\includegraphics[width=\linewidth]{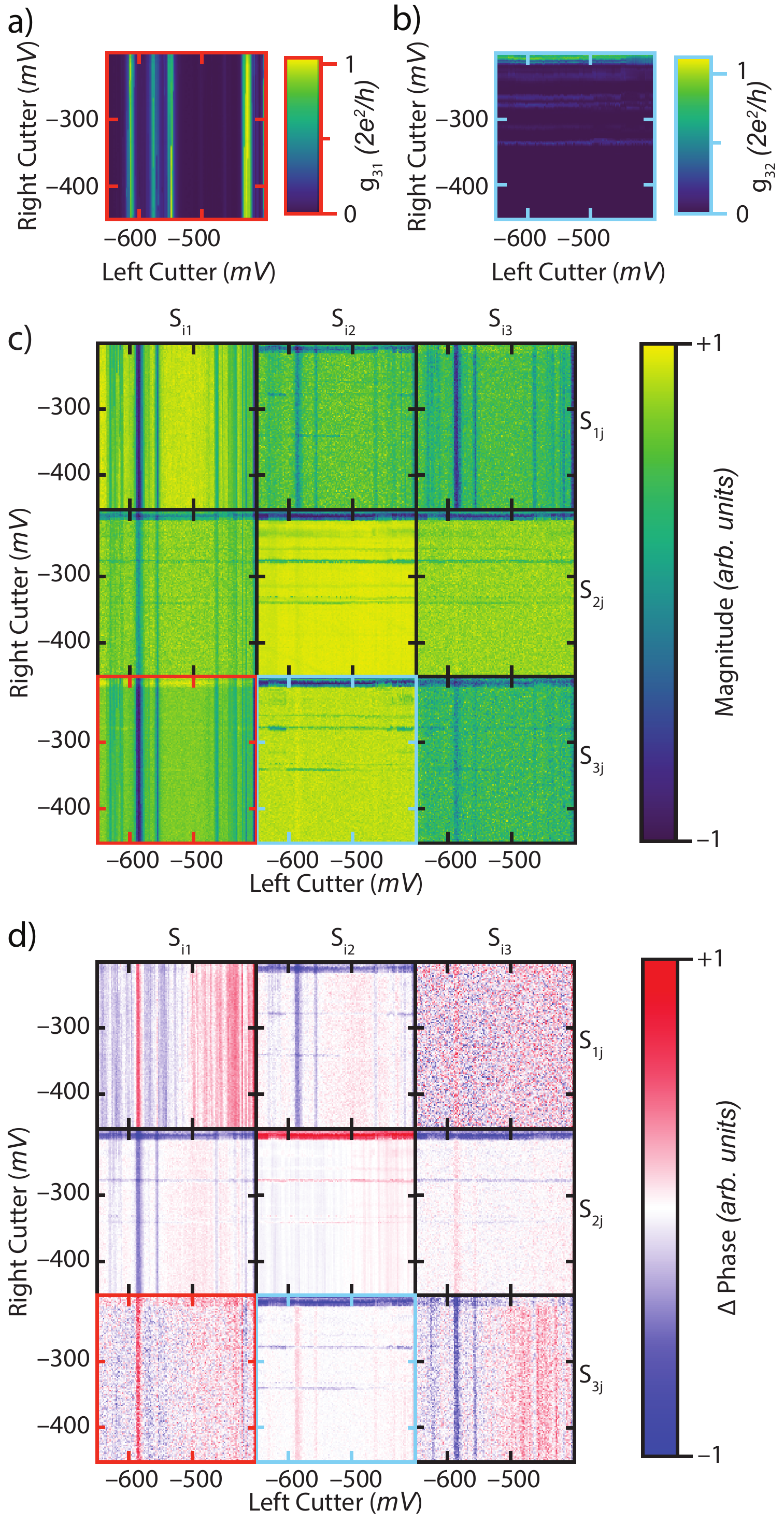}
\caption{\label{Fig_Tartan} 
Cutter-gate verse cutter-gate scans for a plunger gate of -1.99~V.
(a) Left to middle  measurement of conductance, $g_{31}$, as a function right and left cutter-gate, with ac excitation for transport applied to the left terminal (port 1).
(b) Right to middle measurement of conductance, $g_{32}$, as a function right and left cutter-gate, with transport excitation applied to the right terminal (port 2).
(c) Magnitude response of $S$-parameters at 3.2~GHz. Parameters involving the left (right) hand side - Port 1 (2) - show sensitivity predominantly to the left (right) cutter. 
(d) Difference in phase response for $S$-parameters, with respect to the median value, at 3.2~GHz. Vertical (horizontal) lines are predominantly present in parameters involving the left (right), port 1 (2). 
}
\end{figure}
In what follows we examine the microwave $S$-parameter approach in comparison to low frequency (dc) transport, beginning with a straightforward measurement of the conductance as a function of left and right cutter gate voltage $V_{LC}$
and $V_{RC}$, as shown in Fig.~\ref{Fig_Tartan}.(a) and \ref{Fig_Tartan}.(b). Here, we measure the local conductances, $g_{13}$ and $g_{23}$, which probe local transport between the left or right normal lead and the middle superconducting contact. As the gates pinch off, the data exhibits peaks and dip features that are typical for these devices and are likely due to charge state resonances associated with the disorder potential. We note that vertical line features appear in the left conductance and horizontal line features in the right when the data is plotted with right gate bias on the vertical axis and left gate bias on the horizontal. When measuring from the left the (vertical) features are strongly dependent on the left cutter gate voltage, but exhibit no dependence on the right cutter gate voltage. The opposite is true for the horizontal features when probing from the right.
Taken together these patterns suggest that vertical features are related to states associated with the left tunnel barrier and horizontal features relate to states near the right barrier. Such patterns are well-known in the quantum dot literature \cite{qdots}. 

Turning to the microwave measurements, Figs.~\ref{Fig_Tartan}.(c) and ~\ref{Fig_Tartan}.(d) show the magnitude and phase of the 9-element $S$-parameter matrix, taken at a frequency of 3.2~GHz and zero magnetic field. Similar data is seen at other frequencies (acquired simultaneously) across the band. We first draw attention to the correspondence between the conductance  measurements $g_{31}$ and $g_{32}$ and the $S$-parameters $S_{31}$ (red box) and $S_{32}$ (light blue box) [refer to the key in Fig.~\ref{fig1}.(d) to aid interpretation]. Given that port 1 is the left terminal and port 2 is the right, it is perhaps not surprising that data for these elements strongly resemble the local dc transport data. We note that $S_{11}$ and $S_{22}$ data, which measure the standard response of a reflectometry setup, also show features related to local transport since, ideally $|S_{11}|^2 \approx 1 - |S_{31}|^2$, given $|S_{21}|^2 \ll |S_{31}|^2$ in these systems \cite{pan2020}. More interesting are the $S$-matrix elements that relate to the non-local response, $S_{12}$ and $S_{21}$. We note that if all of the microwave power was truly shunted to ground by the middle superconducting contact, the non-local elements $S_{21}$ and $S_{12}$ should display negligible signal near zero bias. Instead, we find these data-sets contain both vertical and horizontal features consistent with a measurement that that probes both tunnel barriers in series.
\begin{figure*}
\includegraphics[width=\linewidth]{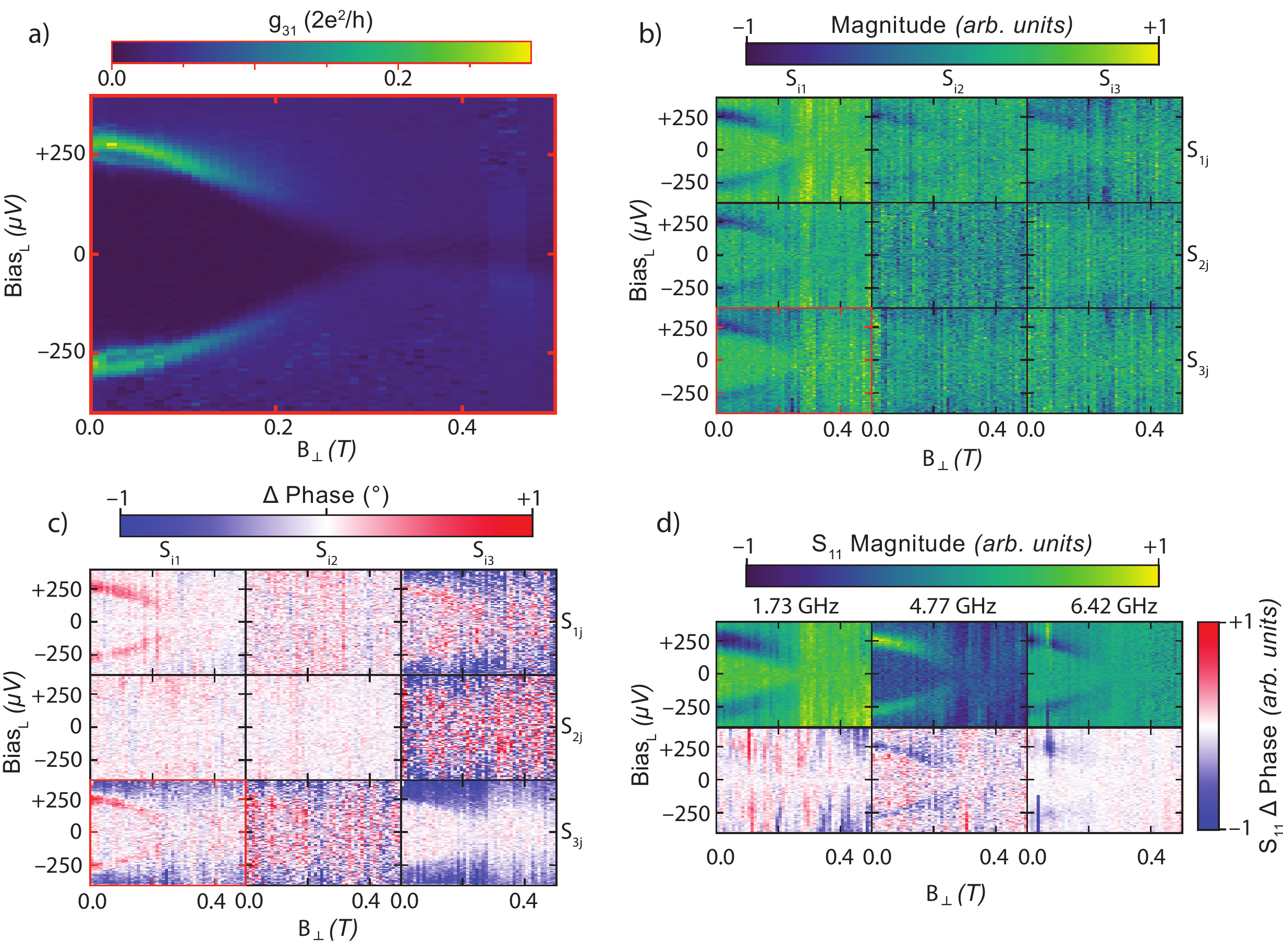}
\caption{\label{Gap_closing} 
Measurements with the magnet field applied in-plane with the wire and perpendicular to the main axis, with bias applied from the left. 
(a) Conductance response $g_{31}$ of the left to middle terminal measured.
(b) Magnitude and 
(c) phase of $S$-parameters, at 4.54~GHz; with the background offset subtracted.
(d) Magnitude and phase of the S$_{11}$ parameter with background offset subtracted, for 1.73, 4.77, and 6.42~GHz.
}
\end{figure*}

Realizing that the different $S$-parameters may simultaneously reveal both the local and non-local response of the device, we test this concept further by performing bias spectroscopy of the superconducting gap as it closes with magnetic field. Here, the field is applied perpendicular to the major axis of the wire with conductance $g_{31}$ again measured between the left normal lead and the middle (grounded) terminal. Raising the tunnel barriers by applying negative voltages to the left and right cutter gates, we perform bias spectroscopy, as shown in Fig.~\ref{Gap_closing}(a). Peaks in the differential conductance associated with the Bardeen-Cooper-Schrieffer (BCS) density of states define the edges of the superconducting gap that closes with increasing field. At zero magnetic field it appears that the induced gap in this device is very similar to the parent gap, $\Delta_{ind} \approx \Delta_{Al}$, consistent with similar transport measurements \cite{Puglia2020}.

The corresponding microwave measurements are shown in Figs.~\ref{Gap_closing}(b) and \ref{Gap_closing}(c) where we again plot the magnitude and phase response of the $S$-parameters, with bias and field [refer to the key in Fig.~\ref{fig1}(d) to aid interpretation]. For comparison with the dc transport data in Fig.~\ref{Gap_closing}(a), we first look to the configuration $S_{31}$ (element with red border), ie, the magnitude of transmitted power that flows from port 1 (left normal lead) to port 3 (middle superconducting contact). In correspondence with transport, clear evidence of a normal-superconductor tunnel barrier manifests as two peaks in the microwave response that can be used to determine the superconducting gap and its dependence on magnetic field. With the dc bias applied to the left normal lead (port 1), we note that the two-peak BCS signature of the gap shows up in all measurements involving port 1, ie, $S_{i1}$ and $S_{1j}$, where $i,j$ = 1-3, in Fig.~\ref{Gap_closing}(b). 

\begin{figure*}
\includegraphics[width=\linewidth]{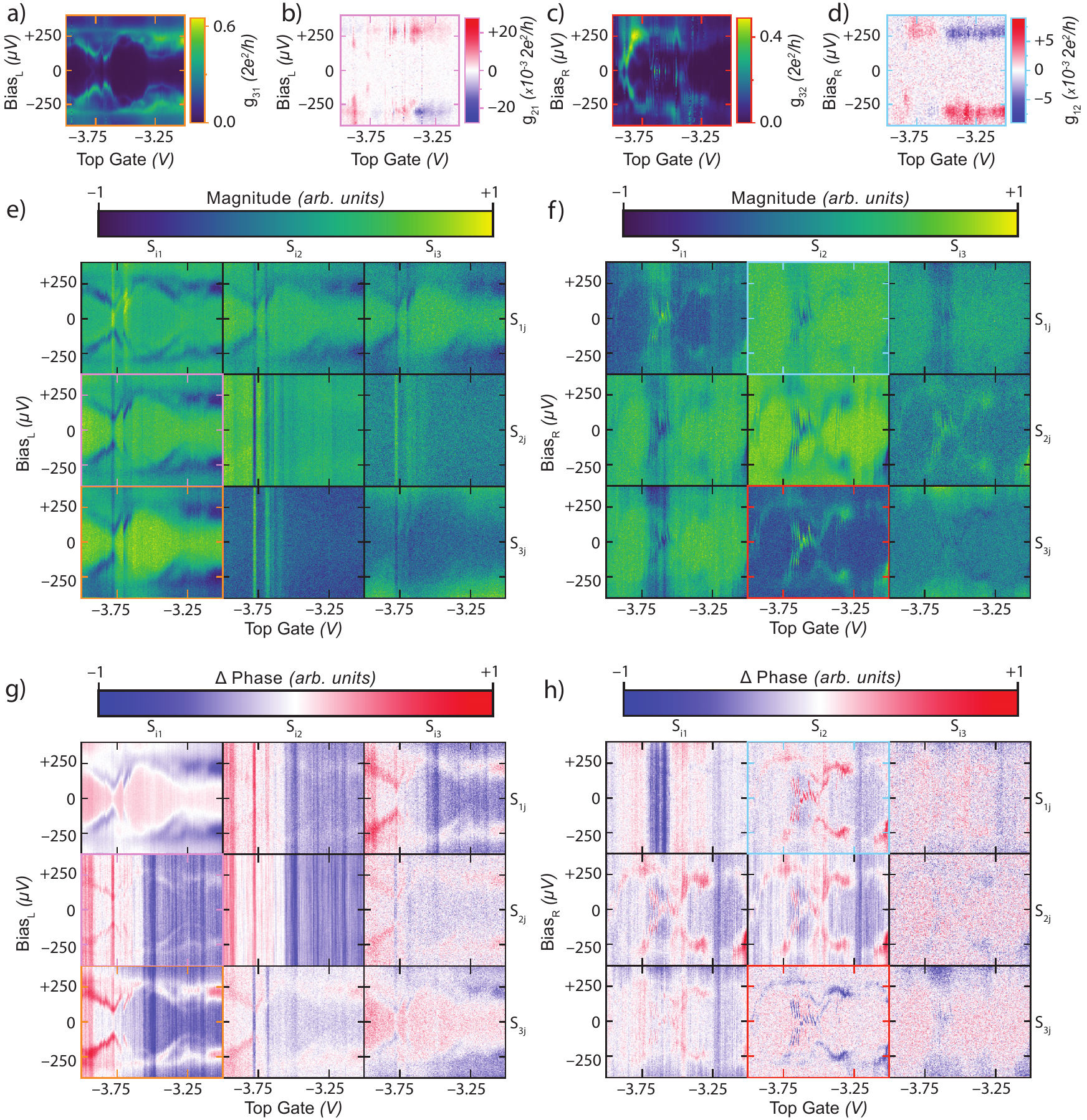}
\caption{\label{Fig_Bias_comp} 
Zero magnetic field sub-gap features measured with dc transport and the microwave technique. Local-conductance measurement 
(a) and non-local conductance 
(b) measured with the bias applied from the left, middle and right terminal grounded. 
Similarly (c) and (d) show the dc transport local and non-local conductance with the device biased from the right.
(e,~g) the magnitude and phase with bias applied from the left, $S$-parameters acquired at 4.795~GHz, similarly
(f,~g) but with the bias applied from the right. 
}
\end{figure*}

More interesting is the phase response shown in Fig.~\ref{Gap_closing}(c). Looking again at $S_{31}$ we note that the peaks at the gap edge result in a positive phase shift (red), and that these appear at a slightly lower bias or energy than the features associated with a negative phase shift (blue) that do not evolve significantly with field. Noticing that the blue features essentially occur only for measurements that involve the middle superconducting terminal (i.e., $S_{i3}$ or $S_{3j}$), we suggest that this negative phase shift (relative to zero bias) may result from the weaker field response of the parent (Al) superconductor [noting again that the field is perpendicular to the long axis of the nanowire]. Further data is needed to test this hypothesis. 

As discussed above, the TRL technique can enable measurements at different frequencies. As an example of this, Fig.~\ref{Gap_closing}(d) shows the local response from the left lead, measured at 1.73~GHz, 4.77~GHz, and 6.42~GHz. The signal at 4.77~GHz appears stronger than at other probe frequencies, perhaps because this frequency is comparable to the tunnel rate. 

Lastly, we examine sub-gap features at zero magnetic field that are frequently observed in these disordered devices when varying gate biases \cite{Puglia2020}. Here we acquire local and non-local conductance together with microwave measurements and compare data sets taken with the dc bias applied on the left or right side of the device with respect to the grounded middle terminal. Starting with dc transport, Figs.~\ref{Fig_Bias_comp}(a) and (b) show the local $g_{31}$ and non-local $g_{21}$ conductance, with the dc bias applied between port 1 (left lead) and port 3 (ground). Conversely, Figs.~\ref{Fig_Bias_comp}(c) and (d) show the corresponding conductance measurements for the case where the dc bias is applied between port 2 (right lead) and port 3. 

Both bias configurations show significant sub-gap structure in the local conductance data, consistent with similar experiments on these devices \cite{Puglia2020,Anselmetti2019,Menard2020}. The sensitivity of these features to the top gate bias suggests they result from Andreev bound states (ABSs) and unintentional quantum dots associated with disorder. We note that these sub-gap features appear distinct when biasing from the left or the right, likely because they originate from separate ABSs. In contrast, the non-local conductance data shown in Figs.~\ref{Fig_Bias_comp}(b) and (d) appears relatively clean, with only a very small region near $\pm$ 260~$\mu$V contributing to the signal. Recalling that the non-local conductance stems from transport in the window between the induced gap and parent superconducting gap, a small signal is not unexpected given $\Delta_{ind} \approx \Delta_{Al}$ at $B$ = 0 in these devices. 
\begin{figure}
\includegraphics[width=8.5cm]{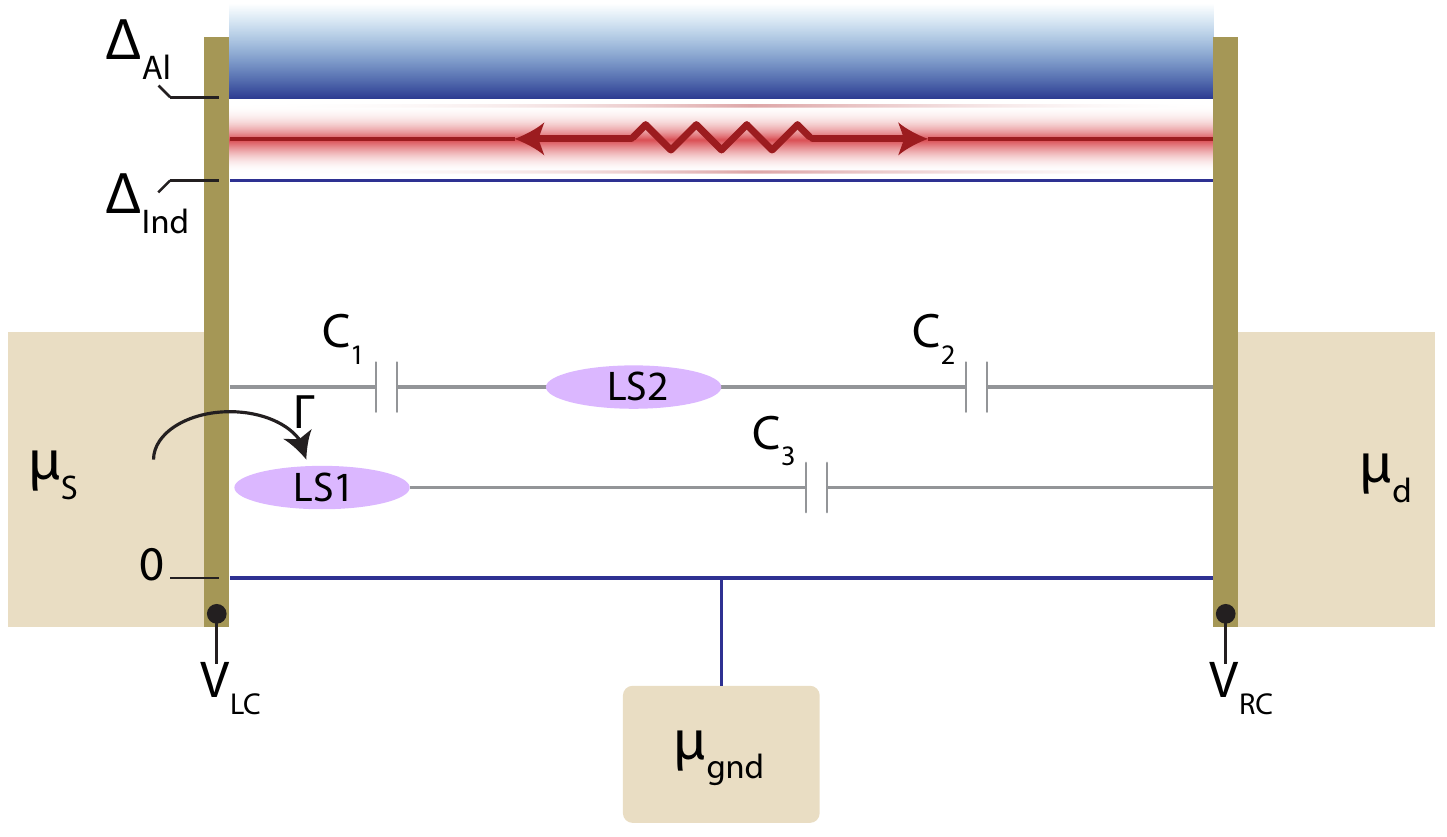}
\centering
\caption{\label{Cartoon} Illustration depicting the energy levels of a three terminal device with localized states (LSs). Here, $\mu_S$, $\mu_d$  and $\mu_{gnd}$ are the chemicals of the terminals, where $\Delta_{Ind}$ and $\Delta_{Al}$ represent the induced and parent superconducting gap of aluminium. Tunnel barriers are defined by gate voltages $V_{LC}$ and $V_{RC}$. In this picture, the microwave response between ports involve contributions from both tunneling and capacitive coupling to localized states. Red arrows indicate non-local transport.}
\end{figure}

Looking now to the corresponding microwave $S$-parameters shown in Figs.~\ref{Fig_Bias_comp}(e) to (h), the picture becomes more complicated. First, we draw attention to the local responses $S_{11}$, $S_{31}$ associated with the left normal lead that is sourcing the dc bias [Figs.~\ref{Fig_Bias_comp}(e) and (g)]. The sub-gap structure observed in the local conductance measurements also shows up here in the local microwave response. Interestingly, the local response from the opposite side with respect to the dc bias shows essentially no signal, ie, $S_{22}, S_{23}$. A consistent picture emerges looking at the data taken with the dc bias applied from the right side [Figs.~\ref{Fig_Bias_comp}(f) and (h)]. Again, sub-gap structure that strongly resembles the local conductance data is observed in the local responses $S_{22}$ and $S_{32}$. Data characterizing the local response from the opposite side ($S_{11}$, $S_{31}$) however, does show some similar sub-gap structure, albeit weaker (particularly weak in the phase response). We discuss the origin of this weaker response below. 

In contrast to the non-local dc transport data, the non-local $S$-parameters $S_{21}$ and $S_{12}$ show significant sub-gap structure that resembles a duplicate of the local (conductance) response, but with a somewhat weaker signal. To explain why $S_{21}$ and $S_{12}$ appear to contain a mixture of the local and non-local conductance response we suggest that these measurements can probe states that are directly accessible via tunneling or alternatively via capacitive pathways. 

We illustrate these mechanisms in Fig.~\ref{Cartoon}, where microwave signals can be coupled from a lead directly to sub-gap localised states via tunnel processes with rate $\Gamma$, or  capacitively via parasitic capacitances $C_{1,2,3}$. These capacitive mechanisms provide a means of coupling ports 1 and 2 and exist in parallel to the non-local conductance window that opens at energies between the induced and parent gaps. Although more complicated to interpret than non-local dc transport, the non-local microwave response may provide additional information by enabling the capacitive probing of `dark' localized  states that are cut off from the leads. 

In closing, we return to the issue of measurement speed, noting that the bandwidth of the microwave technique (0.1 - 1 kHz) presented here exceeds the dc transport bandwidth (10 Hz) by an order of magnitude or more for a comparable signal to noise ratio.  This increase in measurement bandwidth or sensitivity has obvious advantages when searching a large parameter space for features in the data \cite{Pikulin2021, TopoFound}. Beyond the improvement in speed, the additional information provided by measuring the full impedance of these complex devices as a function of frequency is also likely of use in understanding the behaviour of superconductor-semiconductor systems with 3-terminal geometries. 

\section*{Acknowledgements}
The authors gratefully acknowledge the Microsoft HCO team for device fabrication: Paschalis Dalampiras, Agnieszka Telecka, Maren Elisabeth Kloster, Shivendra Upadhyay as well as Emily Toomey and Karl Petersson for important discussions. 
This research was supported by Microsoft Corporation and the Australian Research Council Centre of Excellence for Engineered Quantum Systems (EQUS, No. CE170100009). The authors acknowledge the facilities as well as the scientific and technical assistance of the Research \& Prototype Foundry Core Research Facility at the University of Sydney, part of the Australian National Fabrication Facility.\\

* B.~Harlech-Jones and S.~J.~Waddy contributed equally to this work.

$\dagger$ Corresponding author: David.Reilly@sydney.edu.au

\bibliography{biblio}

\end{document}